\documentclass[twocolumn]{jpsj3}

\title{Growth Dynamics of Photoinduced Domains in Two-Dimensional
Charge-Ordered Conductors Depending on Stabilization Mechanisms
}

\author{
Yasuhiro \textsc{Tanaka$^1$}\thanks{yasuhiro@ims.ac.jp}
and Kenji \textsc{Yonemitsu$^{1,2}$}\thanks{kxy@ims.ac.jp}
}

\inst{
$^1$Institute for Molecular Science, Okazaki, Aichi 444-8585, Japan\\
$^2$Department of Functional Molecular Science, Graduate University for
Advanced Studies, Okazaki, Aichi\ 444-8585, Japan
}

\date{\today}

\abst{
Photoinduced melting of horizontal-stripe charge orders in
quasi-two-dimensional organic 
conductors $\theta$-(BEDT-TTF)$_2$RbZn(SCN)$_4$
[BEDT-TTF=bis(ethylenedithio)tetrathiafulvalene] and
 $\alpha$-(BEDT-TTF)$_2$I$_3$
 is investigated theoretically. By
 numerically solving the time-dependent Schr$\ddot{\rm o}$dinger
 equation, we study the photoinduced dynamics in extended
 Peierls-Hubbard models on anisotropic triangular lattices within the
 Hartree-Fock approximation. The
 melting of the charge order needs more energy for
 $\theta$-(BEDT-TTF)$_2$RbZn(SCN)$_4$ than for
 $\alpha$-(BEDT-TTF)$_2$I$_3$, which
 is a consequence of the larger stabilization energy in
 $\theta$-(BEDT-TTF)$_2$RbZn(SCN)$_4$. After
 local photoexcitation in the charge ordered states, the growth of a
 photoinduced domain shows anisotropy. In
 $\theta$-(BEDT-TTF)$_2$RbZn(SCN)$_4$, the domain hardly expands to the
 direction perpendicular to the horizontal-stripes. This is because all
 the molecules on the hole-rich stripe are rotated in one direction and
 those on the hole-poor stripe in the other direction. They modulate
horizontally connected transfer integrals homogeneously, stabilizing
the charge order stripe by stripe. In $\alpha$-(BEDT-TTF)$_2$I$_3$,
lattice distortions locally stabilize the charge order so that it is
easily weakened by local photoexcitation. The photoinduced
domain indeed expands in the plane. These results are consistent with
recent observation by femtosecond reflection spectroscopy.
}

\kword{photoinduced phase transition, charge order, metal-insulator transition}

\begin{document}
\sloppy
\maketitle

\section{Introduction}
Low-dimensional organic conductors are known to offer a playground for
studying strongly correlated electron systems in which various
interesting ground states emerge\cite{Seo_ChemRev04,Seo_JPSJ06Rev}. The typical
examples are Mott and charge-ordered insulating states, where
electron-electron as well as electron-phonon (e-ph) interactions have
important roles. In such systems, photoirradiation may induce a phase
transition, which has been extensively studied
recently\cite{Nasu_04,Yonemitsu_PhysRep08,Okamoto_PRL06,Iwai_PRL07}
since it will lead to
novel transient phenomena and possible control of various functions of
materials.

The quasi-two-dimensional organic conductors
$\theta$-(BEDT-TTF)$_2$RbZn(SCN)$_4$ (abbreviated as $\theta$-RbZn hereafter)
and $\alpha$-(BEDT-TTF)$_2$I$_3$ (as $\alpha$-I$_3$) are typical compounds
that exhibit a charge order (CO)\cite{Miyagawa_PRB00,Chiba_JPCS01,Takano_JPCS01,Takano_SynMet01}. 
They consist of stacking layers of monovalent anions and donor BEDT-TTF
molecules whose $\pi$-band is 3/4-filled. The photoinduced melting of
the COs in these systems has recently been observed by
using femtosecond reflection spectroscopy\cite{Iwai_PRL07,Iwai_PRB08}. 
It shows a marked difference between their photoinduced dynamics.
For $\alpha$-I$_3$, a semimacroscopic metallic domain is generated,
whereas the CO only locally melts for $\theta$-RbZn. 
In particular, the dynamics in $\alpha$-I$_3$ shows critical slowing
down and strong dependence on the excitation
intensity and temperature\cite{Iwai_PRL07}. 

This clear difference is considered to originate from different roles of
lattice distortions in stabilizing the COs. In fact, the CO transition in
$\theta$-RbZn is a first-order metal-insulator transition with large
structural distortion at $T_c=200$K\cite{Mori_PRB98,M_Watanabe_JPSJ04}. 
On the other hand, $\alpha$-I$_3$ undergoes small lattice distortion at
the CO transition with
$T_c=135$K\cite{Bender_MCLC84,Kakiuchi_JPSJ07}. In both salts,
a horizontal-stripe CO is formed, as confirmed by several
experiments such as X-ray
scattering\cite{M_Watanabe_JPSJ04,Kakiuchi_JPSJ07} and Raman
spectroscopy\cite{Yamamoto_PRB02,Wojciechowski_PRB03}.

Theoretically, CO phenomena have been investigated by using extended
Hubbard models including on-site ($U$) and intersite ($V_{ij}$)
Coulomb
interactions\cite{Seo_JPSJ00,Mckenzie_PRB01,Clay_JPSJ02,Mori_JPSJ03,%
Merino_PRB05,Kaneko_JPSJ06,Watanabe_JPSJ06,Kuroki_JPSJ06,Hotta_PRB06,%
Hotta_JPSJ06,Seo_JPSJ06,Udagawa_PRL07,Tanaka_JPSJ07,Miyashita_PRB07,%
Tanaka_JPSJ08,Tanaka_JPSJ09ERRATA,Miyashita_JPSJ08,Nishimoto_PRB08}.
The stability of various CO patterns in some members of (BEDT-TTF)$_2$X has
been discussed first within the Hartree approximation by considering
realistic band structures of
(BEDT-TTF)$_2$X\cite{Seo_ChemRev04,Seo_JPSJ00}. The horizontal-stripe CO is
shown to be stabilized in $\theta$-RbZn and also in
$\alpha$-I$_3$\cite{Seo_JPSJ00} if the transfer integrals are based on
the low-temperature crystal structures with lattice distortions
implicitly included. The effects of e-ph interactions on the
horizontal COs have been investigated by the Hartree-Fock approximation
and exact
diagonalization\cite{Tanaka_JPSJ07,Tanaka_JPSJ08,Tanaka_JPSJ09ERRATA,Miyashita_PRB07,Miyashita_JPSJ08}. 
Here the transfer integrals are based on the high-temperature crystal
structures if the lattice is undistorted.
The results show that charge frustration due to the Coulomb
interactions on a triangular lattice and lattice distortion relieving
the frustration are essential for $\theta$-RbZn,
while the low-symmetry configuration of transfer integrals is important
for $\alpha$-I$_3$. These theoretical studies consistently interpret the
experimental findings including the large discontinuity at
the CO transition in $\theta$-RbZn and the difference between the
behaviors of the spin degrees of freedom in the two salts below
$T_c$\cite{Mori_PRB98,Rothamael_PRB86}. Therefore, it is
of great importance to investigate the photoinduced melting of the COs
and to compare their dynamics.

In this paper, we investigate the photoinduced melting dynamics in
$\theta$-RbZn and in $\alpha$-I$_3$ using the time-dependent
Schr$\ddot{\rm o}$dinger equation for extended Hubbard models with
Peierls-type e-ph couplings within the Hartree-Fock approximation. It is
found that the CO in $\theta$-RbZn is more stable against
photoexcitation than that in $\alpha$-I$_3$, which is consistent with
the experimental results\cite{Iwai_PRL07}. Their dynamics show different
behaviors owing to different roles of e-ph couplings for the COs
and different underlying crystal structures. In particular, a
photoinduced domain with suppressed CO grows anisotropically in
$\theta$-RbZn, whereas it grows isotropically in $\alpha$-I$_3$. This
suggests that a macroscopic domain is more easily created in
$\alpha$-I$_3$ than in $\theta$-RbZn.
In \S 2, the extended Peierls-Hubbard model is defined. The
numerical method for solving the time-dependent
Schr$\ddot{\rm o}$dinger equation is also given. 
After presenting linear absorption spectra in \S 3, we show photoinduced
dynamics during and after a spatially uniform, oscillating electric
field is introduced in \S 4. The growth of a photoinduced domain after
local melting of the CO is discussed in \S 5. Section 6 is devoted to a summary.

\section{Extended Peierls-Hubbard Model on Triangular Lattice}
In order to describe the horizontal CO with lattice distortion in
$\theta$-RbZn and $\alpha$-I$_3$, we
consider the following extended Peierls-Hubbard
model\cite{Tanaka_JPSJ07,Tanaka_JPSJ08,Tanaka_JPSJ09ERRATA,Miyashita_PRB07,Miyashita_JPSJ08}
\begin{eqnarray}
H = H_{\rm el}+H_{\rm lat}\ ,
\end{eqnarray}
with
\begin{eqnarray}
H_{\rm el}&=&\sum_{\langle ij \rangle \sigma} \left[
(t_{i,j} + \alpha_{i,j} u_{i,j}) 
e^{i(e/\hbar c) \mbox{\boldmath $ \delta $}_{i,j} \cdot \mbox{\boldmath $ A $}(t)}
c^\dagger_{i\sigma} c_{j\sigma} +\mbox{H.c.}
\right] \nonumber \\ 
&+&U\sum_i n_{i\uparrow} n_{i\downarrow}
+\sum_{\langle ij \rangle} V_{i,j} n_i n_j\ ,\\
H_{\rm lat}&=&\sum_{\langle ij \rangle } \frac{K_{i,j}}{2} u_{i,j}^2
+\sum_{\langle ij \rangle } \frac{K_{i,j}}{2\omega_{i,j}^2}
\dot{u}_{i,j}^2\ ,
\end{eqnarray}
where $\langle ij\rangle$ represents the summation over pairs of
neighboring sites, $c^{\dagger}_{i\sigma}(c_{i\sigma})$ denotes the
creation (annihilation) operator for an electron with spin $\sigma$ at
the $i$th site, $n_{i\sigma}=c^{\dagger}_{i\sigma}c_{i\sigma}$, and
$n_{i}=n_{i\uparrow}+n_{i\downarrow}$. The electron density is fixed at 
3/4 filling. The on-site repulsion is denoted by $U$. For the intersite
Coulomb interactions $V_{i,j}$, we consider nearest-neighbor
interactions $V_{c}$ in the vertical
direction and $V_{p}$ in the diagonal direction, as shown in
Fig. 1(a). The e-ph coupling constant, lattice displacement, elastic
constant and bare phonon frequency are denoted by $\alpha_{i,j}$, $u_{i,j}$,
$K_{i,j}$ and $\omega_{i,j}$, respectively. 
We introduce new variables by $y_{i,j}=\alpha_{i,j}u_{i,j}$ and
$s_{i,j}=\alpha_{i,j}^{2}/K_{i,j}$. The notations of the transfer integrals
$t_{i,j}$, $y_{i,j}$ and $s_{i,j}$ are the same as those in
refs. 32 and 33 for both compounds. 

The structures of $\theta$-RbZn and $\alpha$-I$_3$ in the high- and
low-temperature phases are schematically shown in Fig. 1. In the
metallic
phase of $\theta$-RbZn, there are two kinds of transfer integrals $t_p$
and $t_c$ for the diagonal and vertical bonds, respectively, while the CO
transition doubles the unit cell in the $c$-direction and six transfer
integrals appear at low temperatures. On the other hand, the unit cell of
$\alpha$-I$_3$ contains four molecules in both phases. 
According to the X-ray
structural analysis\cite{Kakiuchi_JPSJ07}, sites A and A$^{\prime}$ are
equivalent in the metallic phase owing to inversion symmetry, while the
symmetry breaks below the CO transition temperature. 

\begin{figure}
\includegraphics[height=7.0cm]{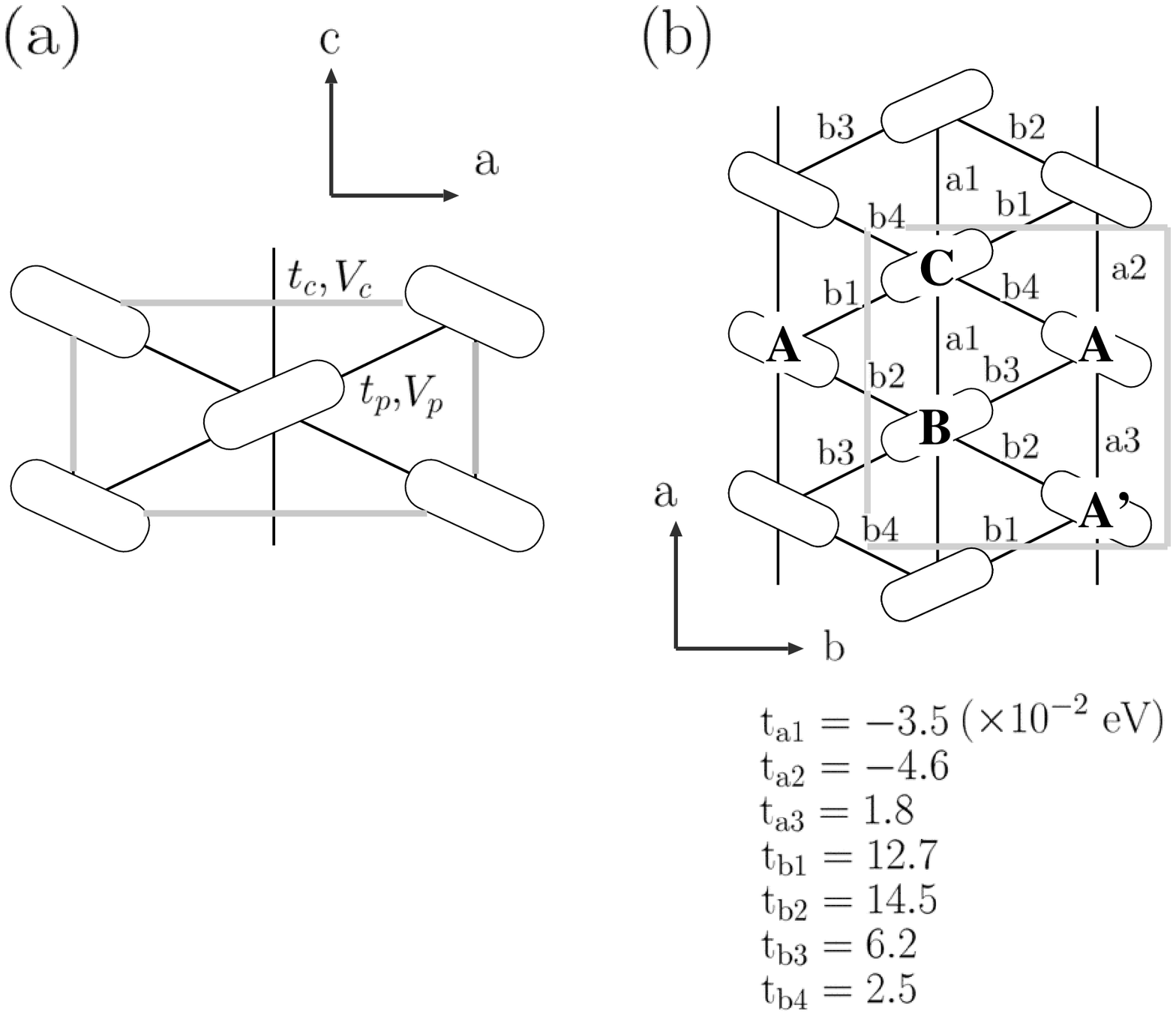}
\includegraphics[height=14.0cm]{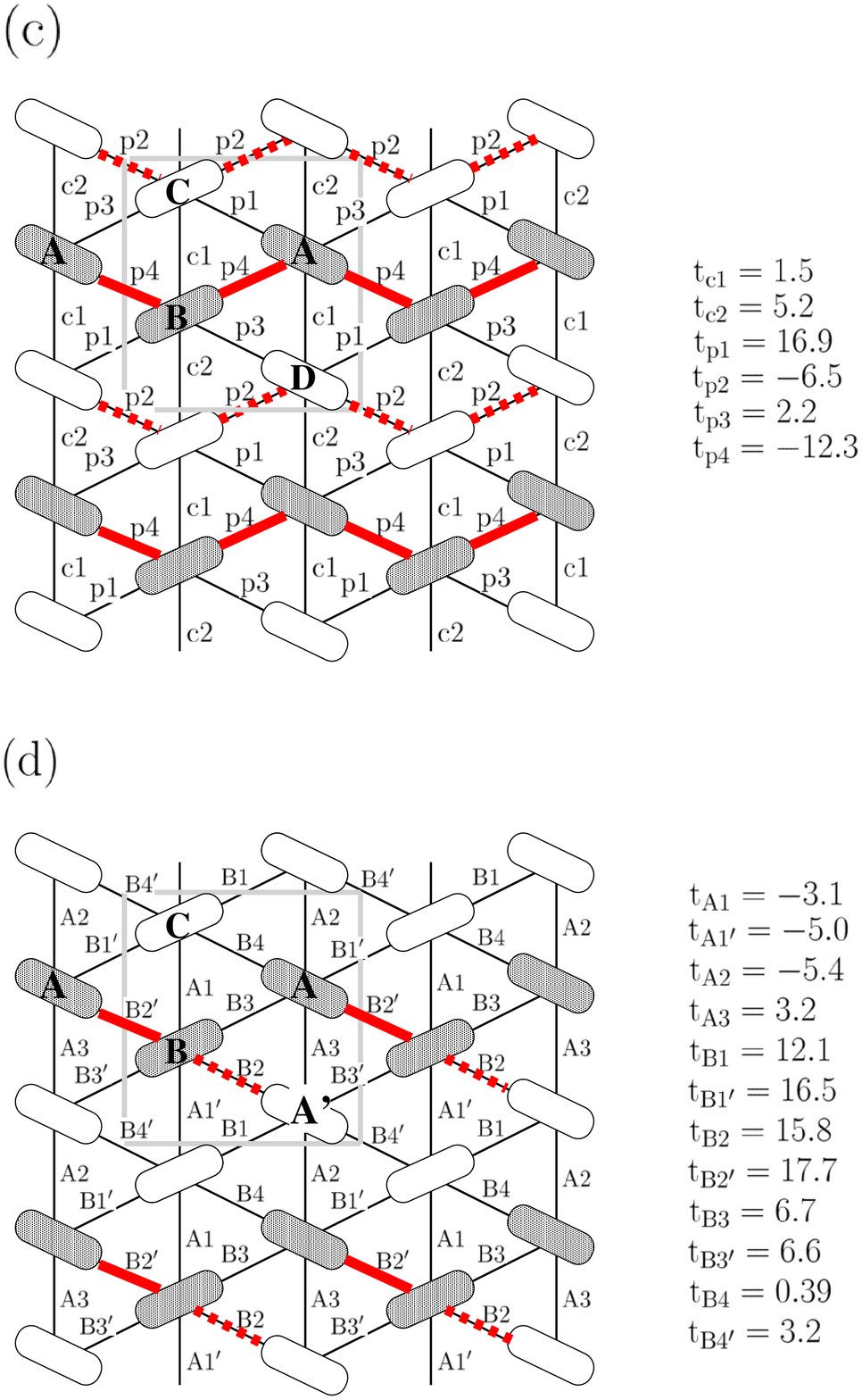}
\caption{(Color online) Schematic representations of the structures of
 (a) $\theta$-(BEDT-TTF)$_2$X in the metallic phase, (b) $\alpha$-I$_3$ in
 the metallic phase\cite{Footnote}, (c) $\theta$-RbZn in the CO phase,
 and (d) $\alpha$-I$_3$ in the CO phase. The gray solid
 lines indicate unit cells. For (b), (c), and (d), the transfer
 integrals estimated by the extended H$\ddot{\rm u}$ckel
 method\cite{M_Watanabe_JPSJ04,Kakiuchi_JPSJ07} are also shown. The red
 or thick solid (dashed)
 lines in (c) and (d) indicate the bonds on which the magnitudes
 of the transfer integrals are increased (decreased) by the distortions
 considered in the present paper.}
\end{figure}

In the previous
studies\cite{Tanaka_JPSJ07,Tanaka_JPSJ08,Tanaka_JPSJ09ERRATA}, we
consider three types of e-ph couplings $s_{c}$, $s_{p1}$, and $s_{\phi}$
for $\theta$-RbZn, and two types of e-ph couplings $s_{b1}$ and $s_{b2}$
for $\alpha$-I$_3$ in order to investigate the roles of all the lattice
distortions on the stability of the COs. For $\theta$-RbZn, the
high-temperature structure has high symmetry, where several patterns of
COs are nearly degenerate because of the charge frustration. 
The lattice distortions relieve the frustration. We found
that $s_{\phi}$, which comes from molecular
rotation\cite{Tanaka_JPSJ07,Miyashita_PRB07}, is the most important for
realizing the horizontal CO. As shown in Fig. 1(c), this distortion
homogeneously modulates the horizontally connected transfer
integrals\cite{Tanaka_JPSJ07,Miyashita_PRB07}, $t_{p2}=t_p-y_{\phi}$ and
$t_{p4}=t_p+y_{\phi}$, by which the CO is
globally stabilized. For $\alpha$-I$_3$, on the other hand, lattice
effects are relatively small and the low-symmetry configuration of
transfer integrals stabilizes the horizontal
CO\cite{Tanaka_JPSJ08,Miyashita_JPSJ08}, as explained below. Charge
disproportionation between sites B and C exists even in the metallic
phase\cite{Kobayashi_JPSJ04,Kobayashi_JPSJ05,Katayama_JPSJ06,Kobayashi_JPSJ07}
because the large and nonequivalent transfer
integrals $t_{b1}$ and $t_{b2}$ form a zigzag
chain\cite{Kakiuchi_JPSJ07}, as seen in Fig. 1(b). Since
$t_{b2}$ is larger than $t_{b1}$ in the chain, site B (C) becomes
hole-rich (hole-poor) so as to gain the kinetic energy. The appearance of
the horizontal CO breaks the equivalence of sites A and
A$^{\prime}$. The e-ph couplings $s_{b1}$ and $s_{b2}$ stabilize this
CO. However, the configuration of large transfer integrals is
not along the hole-rich stripes in contrast to the case of
$\theta$-RbZn. In other words, the CO is locally stabilized in
$\alpha$-I$_3$: the distortion causing $t_{B2^{\prime}}>t_{B2}$
($t_{B1^{\prime}}>t_{B1}$) in Fig. 1(d) locally strengthens the bond
B$2^{\prime}$ (B$1^{\prime}$), which is regarded as the formation
of a local singlet on the bond B$2^{\prime}$. As a consequence, site A
(A$^{\prime}$) becomes hole-rich (hole-poor).
With these facts in mind, in the present paper we consider
only $s_{\phi}$ and $s_{b2}$ as the e-ph coupling in $\theta$-RbZn and
$\alpha$-I$_3$, for simplicity. Because only one phonon mode is taken
into account, we set the bare phonon frequency at 
$\omega_{i,j}=\omega_{\rm ph}$.

The photoexcitation is introduced by the Peierls phase factors of the
transfer integrals in eq. (2). Here \mbox{\boldmath $ \delta $}$_{i,j}$
denotes the position vector from the $i$th site to the $j$th site. The vector
potential \mbox{\boldmath $A$}$(t)$ is given by,
\begin{equation}
\mbox{\boldmath $ A $}(t) = 
-c \int_0^t dt' \mbox{\boldmath $ E $}(t')
\;, \label{eq:vector_potential}
\end{equation}
\begin{equation}
\mbox{\boldmath $ E $}(t) = \mbox{\boldmath $ E $}_\mathrm{ext} 
\theta(t) \theta(T_\mathrm{irr}-t) \sin \omega_\mathrm{ext}t
\;, \label{eq:electric_field}
\end{equation}
where \mbox{\boldmath $ E $}$_\mathrm{ext}$ and $\omega_\mathrm{ext}$
is the amplitude and frequency, respectively, of the time-dependent
electric field \mbox{\boldmath $E$}$(t)$. In eq. (5), we define the
pulse width $T_{\rm irr}=2\pi N_{\rm ext}/\omega_{\rm ext}$ with 
$N_{\rm ext}$ being an integer. $\theta (t)$ is the Heaviside step
function, $\theta (t)=1$ for $t>0$ and $\theta (t)=0$ for $t<0$.
We use $e=1$, $\hbar =1$, and the
intermolecular distance along the $c$-axis of 
$\theta$-RbZn and that along the $a$-axis of 
$\alpha$-I$_3$ as the unit of length. Furthermore, the
intermolecular distance along the $a$-axis of
$\theta$-RbZn and that along the $b$-axis of $\alpha$-I$_3$ are set at
twice the unit length. By these definitions, parameters are given in
units of eV unless otherwise noted. 

The time evolution of the system is calculated as follows. For the
electronic part, we use the time-dependent Schr$\ddot{\rm o}$dinger
equation, 
\begin{equation}
|\psi_{\nu ,\sigma}(t+\Delta t)\rangle = T {\rm
 exp}\Bigl[-i\int^{t+\Delta t}_{t}dt^{\prime}H^{\rm
 HF}_{\rm el}(t^{\prime})\Bigr]|\psi_{\nu ,\sigma}(t)\rangle ,
\end{equation}
where $T$ denotes the time-ordering operator, 
$|\psi_{\nu ,\sigma}(t)\rangle$ the $\nu$-th one-particle state with
spin $\sigma$ at time $t$, and $H_{\rm el}^{\rm HF}$ the electronic part
$H_{\rm el}$ in the Hartree-Fock approximation. 

For the lattice part, we solve the classical equation of motion
by the leapfrog method\cite{Miyashita_JPSJ03}. In the case of
$\theta$-RbZn, we assume that for the $l$th unit cell the transfer
integrals on the two $p4$-bonds 
connected to site B and those on the two $p2$-bonds connected to site C
are given by $t_{p4}(l)=t_p+y_{\phi}(l)$ and
$t_{p2}(l)=t_p-y_{\phi}(l)$, respectively. The equation of motion for
$u_{\phi}(l)$ is then written as,
\begin{equation}
m_{\phi}\frac{d^2 u_{\phi}(l)}{dt^2}=\frac{1}{4}F^{\theta}_{l}(t),
\end{equation}
where we write the force for $\theta$-RbZn as,
\begin{equation}
F^{\theta}_{l}(t) = -\Bigl(\frac{\partial H_{\rm lat}}{\partial u_{\phi}(l)}+
\langle \Psi (t)|\frac{\partial H_{\rm el}^{\rm HF}}{\partial
u_{\phi}(l)}|\Psi (t)\rangle \Bigr).
\end{equation}
Here $m_{\phi}=K_{\phi}/\omega_{\rm ph}^2$ is the phonon mass, and
$|\Psi (t)\rangle$ is the Slater-determinant composed of the
one-particle states $|\psi_{\nu ,\sigma}(t)\rangle$.
In the case of $\alpha$-I$_3$, we set
$t_{B2^{\prime}}(l)=t_{b2}+y_{b2}(l)$ and $t_{B2}(l)=t_{b2}-y_{b2}(l)$.
The equation of motion for $u_{b2}(l)$ thus reads, 
\begin{equation}
m_{b2}\frac{d^2 u_{b2}(l)}{dt^2}=\frac{1}{2}F^{\alpha}_{l}(t),
\end{equation}
where $m_{b2}=K_{b2}/\omega_{\rm ph}^2$ and $F^{\alpha}_{l}(t)$ is
defined as in eq. (8). We note that for
both compounds the initial state is the horizontal CO state where the
lattice displacements do not depend on
$l$\cite{Tanaka_JPSJ07,Miyashita_PRB07,Tanaka_JPSJ08,Tanaka_JPSJ09ERRATA,Miyashita_JPSJ08}.
When the uniform
time-dependent electric field is applied, their response is still
uniform. However, they become nonuniform when the electric field is
applied locally as in \S 5.

In the actual calculations, we first obtain 
$F^{\theta}_{l}(t)$ or $F^{\alpha}_{l}(t)$ by using the wave function
$|\psi_{\nu,\sigma}(t)\rangle$. 
Then, $y_{\phi}$ or $y_{b2}$ at $t+\Delta t$ is calculated, which is
accurate to the order of $(\Delta t)^2$. The wave function 
$|\psi_{\nu,\sigma}(t+\Delta t)\rangle$ is obtained from eq. (6) by,
\begin{equation}
|\psi_{\nu ,\sigma}(t+\Delta t)\rangle \simeq {\rm
 exp}\Bigl[-i\Delta t H^{\rm HF}_{\rm el}(t+\frac{1}{2}\Delta
 t)\Bigr]|\psi_{\nu ,\sigma}(t)\rangle ,
\end{equation}
whose error is of the order of $(\Delta t)^{3}$.
In the above equation, we use $H^{\rm HF}_{\rm el}(t+\frac{1}{2}\Delta t)=
[H^{\rm HF}_{\rm el}(t)+H^{\rm HF}_{\rm el}(t+\Delta t)]/2$ where 
$H^{\rm HF}_{\rm el}(t+\Delta t)$ is constructed 
so that it is accurate to the first order of $\Delta t$. 
The exponential operator is expanded with time
slice $\Delta t =0.01$ until the norm of the wave function becomes unity
with sufficient accuracy. The obtained $|\psi_{\nu ,\sigma}(t+\Delta
t)\rangle$ has an error of the order of $(\Delta t)^3$.

In this paper, the phonon frequency $\omega_{\rm ph}=0.01$ is used for
both compounds. We set $t_p=0.1$ and $t_c=-0.04$ for $\theta$-RbZn. For
the transfer integrals in $\alpha$-I$_3$, we use the values shown in
Fig. 1(b). The system size is fixed at $12\times 12$ sites and the
periodic boundary condition is imposed. The
interaction strengths and the e-ph coupling constants are chosen at
$U=0.7$, $V_c=0.4U$, $V_p/V_c=0.6$, $s_{\phi}=0.1$, and $s_{b2}=0.07$
throughout the study. These parameters give the horizontal CO ground
states for both compounds\cite{Tanaka_JPSJ07,Tanaka_JPSJ08}.

\section{Linear Absorption Spectra}
\begin{figure}
\includegraphics[height=6cm]{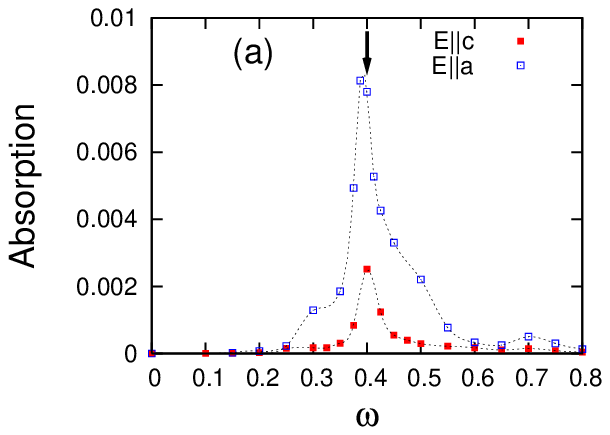}
\includegraphics[height=6cm]{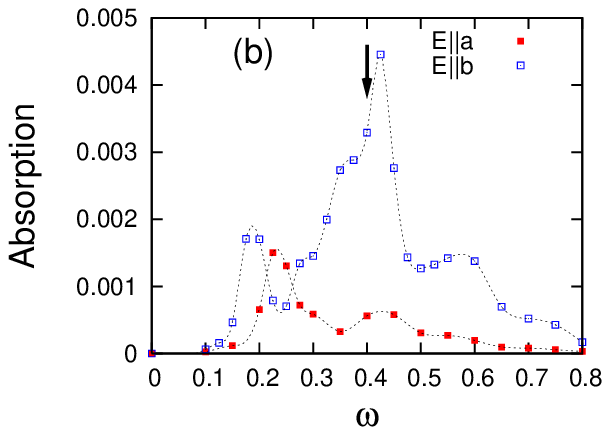}
\caption{(Color online) Linear absorption spectra obtained by the
 time-dependent Hartree-Fock approximation for 12$ \times $12-site
 system with $ U $=0.7, $V_c/U$=0.4, and $V_p/V_c$=0.6 for (a) 
$\theta $-RbZn ($s_{\phi}$=0.1) and (b) $\alpha$-I$_3$ ($s_{b2}$=0.07)
 with different polarizations as indicated. The peak-broadening
 parameter  $\gamma$ is set at 0.02. \label{fig:HF_absorption}}
\end{figure}

First, we discuss linear absorption spectra, which correspond to
optical conductivity spectra. For this purpose, we
replace the electric field in eq. (5) by the one with a damping factor, 
\begin{equation}
\mbox{\boldmath $E$}(t) = \mbox{\boldmath $E$}_\mathrm{ext} 
\theta(t) \exp(-\gamma t) \sin
\omega_\mathrm{ext}t\ , \label{eq:decaying_field}
\end{equation}
and calculate the increment in the total energy due to 
$\mbox{\boldmath $E$}(t)$
with small $\mid \mbox{\boldmath $E$}_\mathrm{ext}\mid$ at sufficiently large
$t$ so that the total energy converges. We use 
$\mid \mbox{\boldmath $E$}_\mathrm{ext}\mid $=0.002 and $\gamma $=0.02
to obtain results shown in Fig. 2. In both cases of $\theta$-RbZn
and $\alpha$-I$_3$, the absorption is larger for the
polarization parallel to the stripes, i.e., along the {\it a}- and {\it
b}-axes, respectively, than for the polarization
perpendicular to them, i.e., along the {\it c}- and {\it a}-axes,
respectively. This is
because the transfer integrals on the diagonal bonds are larger than those
on the vertical bonds. The absorption spectra with polarization parallel to
the stripes have similar structures with resonance peaks located at
around $\omega =0.4$ for both compounds. Therefore, in the following we
show results obtained by the electric field with $\omega_{\rm ext}=0.4$
and polarization parallel to the stripes. 

\section{Photoinduced Melting Dynamics}
Next, we discuss photoinduced melting dynamics. Figure 3 shows the
time evolution of the hole densities $2-\langle n_i\rangle$ and the lattice
distortions during and after the photoexcitation with $N_{\rm ext}=15$
($T_{\rm irr}=236$ corresponding to 155 fs) which is comparable to the
experimental pulse width\cite{Iwai_PRL07}. 
\begin{figure}
\includegraphics[height=16cm]{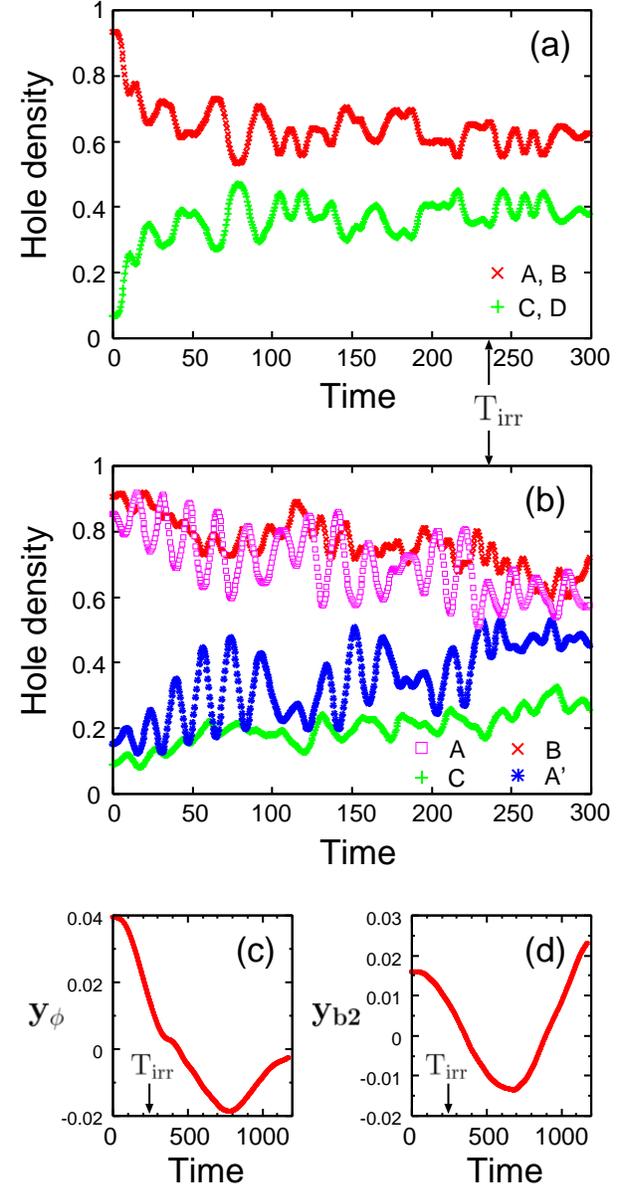}
\caption{(Color online) Time dependence of hole densities during
 ($t<T_{\rm irr}$) and after ($t>T_{\rm irr}$) photoexcitation along the
 stripes for (a) $\theta$-RbZn ($\omega_{\rm ext}$=0.4, 
$\mid \mbox{\boldmath $E$}_{\rm ext}\mid =0.25$) and (b) $\alpha$-I$_3$
 ($\omega_\mathrm{ext}$=0.4,  
$\mid \mbox{\boldmath $E$}_{\rm ext}\mid =0.04$). The time dependence of 
modulations in transfer integrals is also shown in (c) and (d),
 respectively, on a longer timescale.}
\end{figure}
The unit of time is $({\rm eV})^{-1}$, so that $t=1520$ corresponds to
1 ps.
For both salts, the COs are weakened by photoexcitation,
as seen in Figs. 3(a) and 3(b). The amplitude of the
electric field is chosen at 
$\mid \mbox{\boldmath $E$}_{\rm ext}\mid =0.25$ for
$\theta$-RbZn and $\mid \mbox{\boldmath $E$}_{\rm ext}\mid =0.04$
for $\alpha$-I$_3$,
which is near the critical value for melting the CO. The increment in
the total energy per site $\Delta E$ is 0.098 and 0.042, respectively. 

For $\theta$-RbZn, there are two distinct hole densities at 
$t=0$ since each stripe consists of sites with equal hole
densities. This is also true at $t>0$ if the polarization of 
$\mbox{{\boldmath $E$}}(t)$ is parallel to the stripes as in Fig. 3(a),
whereas the hole densities on
the four sites in the unit cell become different when the polarization is
perpendicular to the stripes. For $\alpha$-I$_3$, on the other hand, the
four sites are already distinct at $t=0$ because of the low symmetry of
the crystal structure.

As for the time dependence of the hole densities, resonantly excited
 dynamics is clearly seen for $\alpha$-I$_3$ especially in the
 oscillation of the hole densities on sites A and A$^{\prime}$, the
 periods of which are nearly equal to that of the electric field 
$2\pi/\omega_{\rm ext}$ being 15.7. For $\theta$-RbZn,
 the charge dynamics is more complex than that for $\alpha$-I$_3$
 although $\omega_{\rm ext}$ is near the resonance peak of the linear
 absorption spectrum. Such a difference is also visible in results
 obtained by the exact many-electron wave functions on small
 clusters of 12 sites\cite{Miyashita_unpub}, where the origin of the
different behaviors is interpreted with the distribution of energy
levels above the ground state, which is denser in $\theta$-RbZn than 
in $\alpha$-I$_3$. This is because the high symmetry of the undistorted
 structure in $\theta$-RbZn causes charge frustration. 
Therefore, a larger number of excited states are nearly degenerate and
 involved in the photoinduced dynamics in $\theta$-RbZn, which results
 in complex charge dynamics.

The time dependence of the lattice distortions $y_{\phi}$ and $y_{b2}$
is shown in Figs. 3(c) and 3(d), respectively. Since the COs are
almost completely destroyed in the present parameters, the displacements
become zero at some $t$ although they oscillate with periods longer than
$2\pi/\omega_{\rm ph}$ by phonon softening, which are much longer than
the timescales of charge-transfer excitations. For $\omega_{\rm
ph}=0.01$, the oscillation period is
about 1 ps for both compounds, which is comparable to the experimentally
observed value of 0.7 ps (48 cm$^{-1}$) in the time evolutions of
reflectivity spectra\cite{Iwai_PRL07}.

\begin{figure}
\includegraphics[height=16cm]{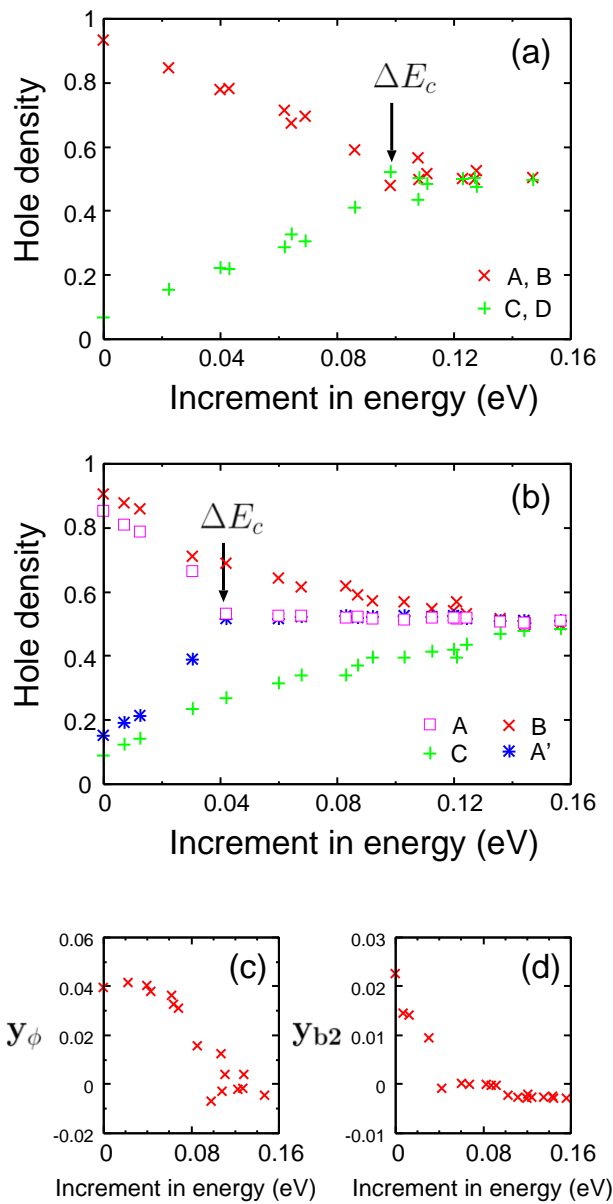}
\caption{(Color online) Time-averages of hole densities after
 photoexcitation along the stripes for (a) $\theta$-RbZn 
($\omega_{\rm ext}$=0.4) and (b) $\alpha$-I$_3$
 ($\omega_{\rm ext}$=0.4), as a function of increment in total energy
 per site $\Delta E$. The time-averages of modulations in transfer
 integrals $y_{\phi}$ and $y_{b2}$ are shown in (c) and (d),
 respectively.}
\end{figure}

In order to compare the efficiency of photoinduced melting for
$\theta$-RbZn and that for $\alpha$-I$_3$, we show
time-averaged hole densities as a function of $\Delta E$ in Figs. 4(a)
and 4(b). Here the averages are taken after $t=T_{\rm irr}$ over the
period of $T_{\rm irr}=2\pi N_{\rm ext}/\omega_{\rm ext}<t<10\pi 
N_{\rm ext}/\omega_{\rm ext}$. 
There is a critical value $\Delta E_c$ above which
the horizontal CO disappears. The value for $\theta$-RbZn, 
$\Delta E_c = 0.098$, is larger than that for $\alpha$-I$_3$, 
$\Delta E_c = 0.042$. This is a consequence of the larger
lattice stabilization energy for the horizontal CO in $\theta$-RbZn than
that in
$\alpha$-I$_3$\cite{Tanaka_JPSJ07,Tanaka_JPSJ08,Tanaka_JPSJ09ERRATA,Miyashita_PRB07,Miyashita_JPSJ08}. 
We notice that this inequality generally holds for any excitation
frequency and for any polarization (not shown). The robustness of the CO
in $\theta$-RbZn compared to that in
$\alpha$-I$_3$ against the photoirradiation has been observed in the
femtosecond spectroscopy\cite{Iwai_PRL07}. The efficiency of the
photoinduced metallic state by $\omega_{\rm ext}=0.89$ eV is evaluated as
100 ($\theta$-RbZn) and 250 ($\alpha$-I$_3$) molecules/photon.
The values of $\Delta E_c$ in
our calculations are, however, one order of magnitude
larger than the experimentally estimated ones. This may be due to the
Hartree-Fock approximation, since $\Delta E_c$ becomes smaller in our
recent study based on the exact many-electron wave
functions\cite{Miyashita_unpub}. Effects of thermal fluctuations and the
experimental estimation itself, which generally contains an error
depending on the estimation of the penetration depth, are also possible
reasons for the discrepancy.

For $\theta$-RbZn, the time-averaged hole distribution
becomes uniform above $\Delta E_c$. For $\alpha$-I$_3$, on
the other hand, the averaged hole densities on sites A and A$^{\prime}$
become equal above $\Delta E_c$ while the charge
disproportionation on sites B and C remains for larger values of $\Delta
E$. This is reminiscent of the fact
that the charge disproportionation exists even in the high-temperature
metallic phase of $\alpha$-I$_3$ because of low-symmetry configuration
of transfer
integrals\cite{Kobayashi_JPSJ04,Kobayashi_JPSJ05,Katayama_JPSJ06,Kobayashi_JPSJ07}.  
If we increase $\Delta E$ further, the hole densities on the four sites
finally merge into $0.5$.

In Figs. 4(c) and 4(d), we show the averaged lattice displacements which
indicate that both $y_{\phi}$ and $y_{b2}$ vanish at around 
$\Delta E_c$. This is reasonable because the $y_{\phi}$ distortion is
essential for stabilizing the horizontal CO in $\theta$-RbZn, while 
the $y_{b2}$ distortion breaks the inversion symmetry that guarantees the
equivalence of hole densities on sites A and A$^{\prime}$ in
$\alpha$-I$_3$. 

\section{Domain Growth after Local Photoexcitation}
Recall that the mechanisms for stabilizing the COs by lattice distortions
are different in the two salts. In $\theta$-RbZn, the whole hole-rich
(hole-poor) stripe is stabilized by strengthening (weakening) the
horizontally connected p4 (p2) bonds. In $\alpha$-I$_3$, the
metallic phase without lattice distortion already possesses a hole-rich
site B and a hole-poor site C by the relation $t_{b2}>t_{b1}$. The
hole-rich site A and the hole-poor site ${\rm A^{\prime}}$ bridged by
the site B are locally stabilized by strengthening the bond
B2$^{\prime}$ and weakening the bond B2. Thus, local photoexcitations
would easily weaken the CO in $\alpha$-I$_3$, but the CO in
$\theta$-RbZn would be robust.

Then, we investigate the growth of photoinduced domains with weakened CO
after local photoexcitation, by modifying the model in eq. (2) through
the introduction of the Peierls phase factors only
on the bonds that connect four sites within a unit cell of the 
$12\times 12$-site system. The phase factor is set at unity on all
the other bonds. The time evolution is calculated by the method
explained in \S 2. In Fig. 5, we show snapshots of the hole densities
at $t=200$, $400$ and $600$, from which those at $t=0$ are subtracted, 
$\langle \Psi (t)|n_i|\Psi (t)\rangle -\langle \Psi (0)|n_i|\Psi
(0)\rangle$, during and after the local photoexcitation with
$\omega_{\rm ext}=0.4$ and 
$N_{\rm ext}=15\ (T_{\rm irr}=236)$. The
polarization is parallel to the stripes. In the figures, the bonds on
which the photoexcitation is applied are located at (7.5,7), (7.5,7.5),
and (7.5,8). Here, the location of the bond between the sites at
$(i_x,i_y)$ and $(j_x,j_y)$ is denoted by
$(\frac{i_x+j_x}{2},\frac{i_y+j_y}{2})$. 
For $\theta$-RbZn ($\alpha$-I$_3$), these bonds correspond
to B-D (B-A$^{\prime}$), B-A, and C-A or p3 (B2), p4 (b3), and p1 (b4),
respectively, in Fig. 1. In order to compare the growth dynamics, we set 
$\mid \mbox{\boldmath $E$}_{\rm ext}\mid =0.6$ for
$\theta$-RbZn and $\mid \mbox{\boldmath $E$}_{\rm ext}\mid =0.32$ for
$\alpha$-I$_3$, which lead to the ratio $\Delta E/\Delta E_c\sim 0.4$
for both compounds. Here $\Delta E_c$ is the value obtained in \S 4 by
the uniform photoexcitation, whereas $\Delta E$ is
the increment in the total energy per site after the local
photoexcitation.

\begin{figure}
\includegraphics[height=10cm]{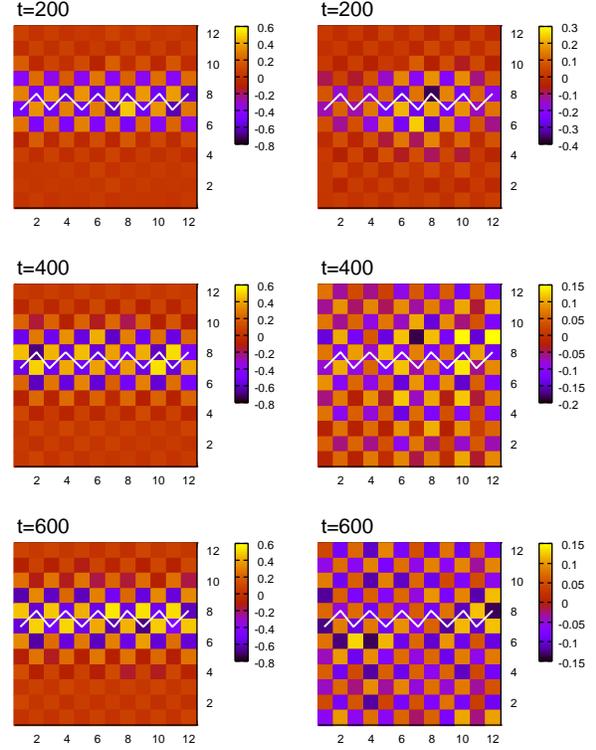}
\caption{(Color online) Hole densities at $t=200, 400$ and $600$
from top to bottom, from which those at $t=0$ are subtracted, for
 $\omega_{\rm ext}=0.4$ and $N_{\rm ext}=15$. The left and right panels
 show $\theta$-RbZn for 
$\mid \mbox{\boldmath $E$}_{\rm ext}\mid =0.6$ and $\alpha$-I$_3$ for 
$\mid \mbox{\boldmath $E$}_{\rm ext}\mid =0.32$, respectively, where the
 squares indicate sites and the white lines indicate hole-rich stripes. The
 photoexcitation with polarization parallel to the stripes is introduced
 only on the bonds (7.5,7), (7.5,7.5), and (7.5,8), which correspond to B-D
 (B-A$^{\prime}$), B-A, and C-A or p3 (B2), p4 (b3), and p1 (b4) 
for $\theta$-RbZn ($\alpha$-I$_3$), respectively.}
\end{figure}

In Fig. 5, the growth of the domain shows
anisotropy, which is pronounced in $\theta$-RbZn. Namely, the growth of
the domain is faster to the direction parallel to the stripes than to
that perpendicular to the
stripes for both compounds. This is caused by the anisotropy in the
transfer integrals which are larger for the $p$- and
$b$-bonds than for the $c$- and $a$-bonds in $\theta$-RbZn and
$\alpha$-I$_3$, respectively. 
\begin{figure}
\includegraphics[height=5cm]{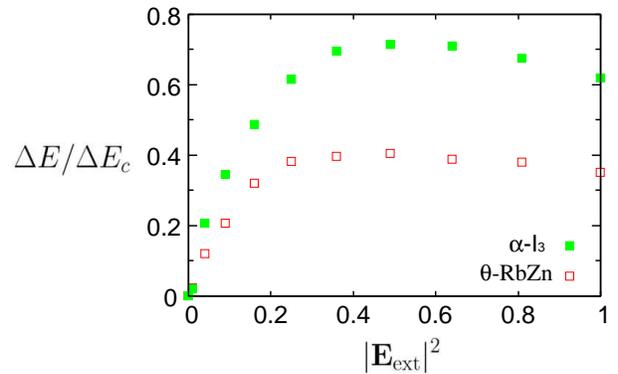}
\caption{(Color online) $\Delta E/\Delta E_c$ as a function of 
$\mid \mbox{\boldmath $E$}_\mathrm{ext}\mid^2 $ for $\theta$-RbZn and
 $\alpha$-I$_3$ after local photoexcitation with polarization parallel
 to the stripes, $\omega_{\rm ext}=0.4$, and $N_{\rm ext}=15$.}
\end{figure}
\begin{figure}
\includegraphics[height=11cm]{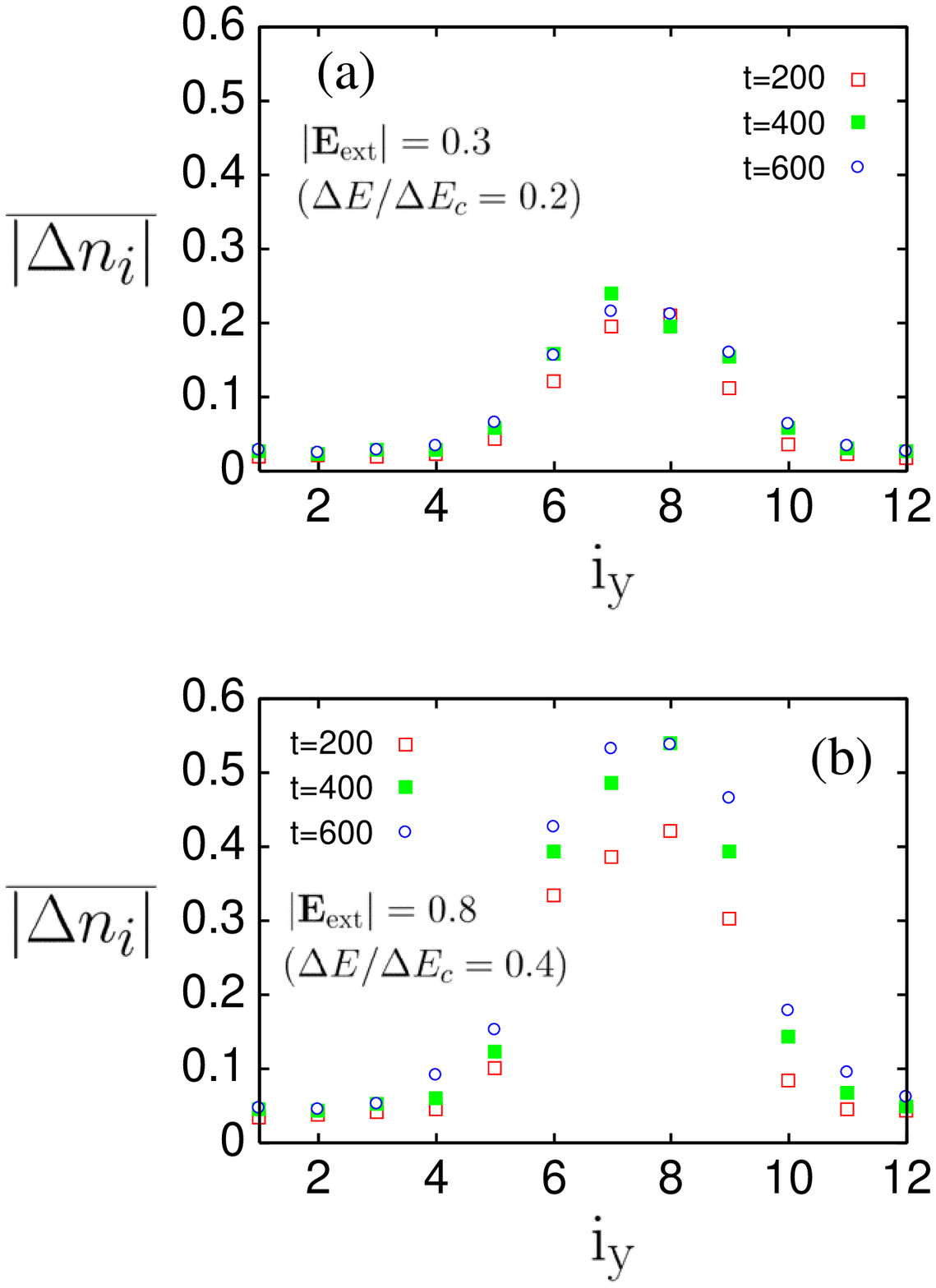}
\caption{(Color online) The $i_y$-dependence of $\overline{|\Delta n_i|}$ at
 $t=200,\ 400$, and 600 for (a) 
$\mid \mbox{\boldmath $E$}_\mathrm{ext}\mid =0.3$ and (b) 
$\mid \mbox{\boldmath $E$}_\mathrm{ext}\mid =0.8$ in the case of
 $\theta$-RbZn, where $i_y$ is the coordinate along the $c$-axis. The
 photoexcitation is along the stripes, $\omega_{\rm ext}=0.4$, and
 $N_{\rm ext}=15$.}
\end{figure}
\begin{figure}
\includegraphics[height=11cm]{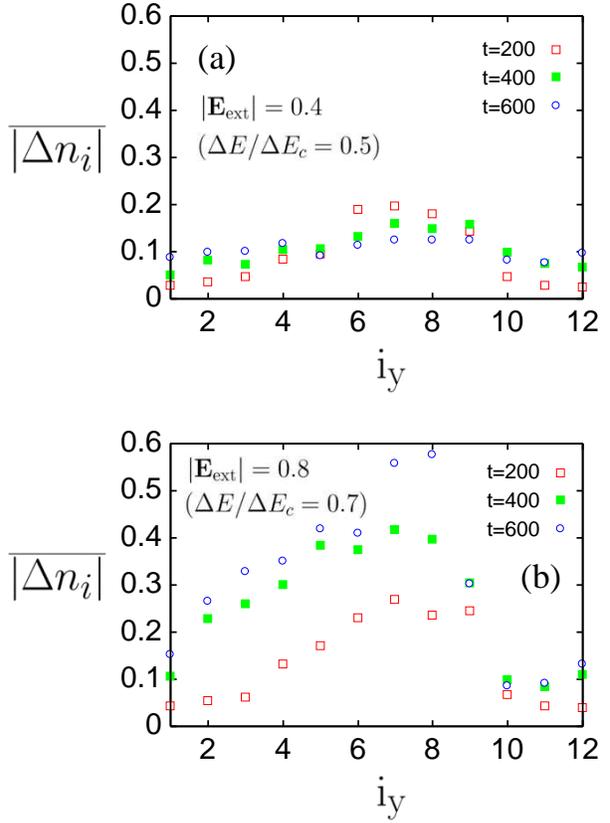}
\caption{(Color online) The $i_y$-dependence of $\overline{|\Delta n_i|}$ at
 $t=200,\ 400$, and 600 for (a) 
$\mid \mbox{\boldmath $E$}_\mathrm{ext}\mid =0.4$ and (b) 
$\mid \mbox{\boldmath $E$}_\mathrm{ext}\mid =0.8$ in the case of
 $\alpha$-I$_3$, where $i_y$ is the coordinate along the
 $a$-axis. The photoexcitation is along the stripes, $\omega_{\rm
 ext}=0.4$, and $N_{\rm ext}=15$.}
\end{figure}
For $\theta$-RbZn, the hole densities largely deviate from
those at $t=0$ only on the horizontal
rows near the four photoexcited sites. The amount of change reaching 0.5
means that the CO is completely destroyed in this region. In the
vertical direction, however, the growth is strongly suppressed. For
$\alpha$-I$_3$, the photoinduced domain expands over the whole system at
$t=600$ although the largest amount
of change in the hole densities is smaller than that in
$\theta$-RbZn. 

In order to analyze the domain growth more quantitatively, we show
$\Delta E/\Delta E_c$ as a function of 
$\mid \mbox{\boldmath $E$}_\mathrm{ext}\mid^2$ in
Fig. 6. 
For both compounds, $\Delta E/\Delta E_c$ has a maximum
value as explained in Appendix, which is below unity. Thus, the CO is
not globally destroyed by
the local photoexcitation with small pulse width. The maximum value is
larger for $\alpha$-I$_3$ than for $\theta$-RbZn, which means that it is
more difficult to weaken the CO in $\theta$-RbZn by the local
photoexcitation with this pulse width. 

The absolute values of the differences between the hole
densities at $t=0$ and those at $t=200,\ 400$ and 600 are averaged over
the direction parallel to the stripes and denoted by 
$\overline{|\Delta n_i|}$ :
\begin{equation}
\overline{|\Delta n_i|}=\frac{1}{L_x}\sum_{i_x}|\langle \Psi
 (t)|n_i|\Psi (t)\rangle -\langle \Psi (0)|n_i|\Psi (0)\rangle |, 
\end{equation}
where $L_x=12$ is the number of sites along the $a$- and $b$-axes for
$\theta$-RbZn and $\alpha$-I$_3$, respectively.
Thus, we can measure the growth of the photoinduced domain to the direction
perpendicular to the stripes. The values $\overline{|\Delta n_i|}$ are
plotted as a function of $i_y$ in Fig. 7
for $\theta$-RbZn and in Fig. 8 for $\alpha$-I$_3$, where $i_y$ is the
coordinate along the $c$- and $a$-axes (which are perpendicular to the
stripes) for $\theta$-RbZn and $\alpha$-I$_3$, respectively, as used in
Fig. 5. Although the charge order on the stripes containing the
photoexcited bonds (7.5,7), (7.5,7.5), and (7.5,8) is weakened and the
quantities $\overline{|\Delta n_i|}$ at $i_y=7$ and 8 are increased by
increasing $\mid \mbox{\boldmath $E$}_\mathrm{ext}\mid$ from 0.3 to 0.8
for $\theta$-RbZn, the photoinduced domain remains localized near
$i_y=7$ and 8. In other words, the photoinduced domain hardly grows to
the direction perpendicular to the stripes. 
The influence of the local photoexcitation is confined in a region
near the photoexcited bonds. This property prevents 
$\Delta E/\Delta E_c$ from becoming large beyond the maximum value, as
shown in Fig. 6.
For $\alpha$-I$_3$, the domain growth is qualitatively different from
that in $\theta$-RbZn. The photoinduced
domain expands to the vertical direction as shown in Fig. 8(a) for
$\mid \mbox{\boldmath $E$}_\mathrm{ext}\mid =0.4$, although the values
of $\overline{|\Delta n_i|}$ are relatively small. When we increase 
$\mid \mbox{\boldmath $E$}_\mathrm{ext}\mid$, a large region of
suppressed CO (i.e., large $\overline{|\Delta n_i|}$) is created as
shown in Fig. 8(b). It is noted that the $i_y$-dependence of 
$\overline{|\Delta n_i|}$ is asymmetric for $\alpha$-I$_3$. 
The photoinduced domain expands to the direction of decreasing $i_y$
more easily than to the direction of increasing $i_y$.
This feature is generally observed irrespective of the choice
of photoexcited bonds because it reflects the asymmetry of the charge
distribution that allows the ferroelectricity\cite{Yamamoto_JPSJ08}.
Let us focus a region near the bond ${\rm B2}^{\prime}$ connecting sites
A and B that has the largest
transfer integral $t_{\rm B2^{\prime}}$ (see Fig. 1). The transfer
integrals $t_{b1}$ and $t_{b4}$ connecting sites A and C are
different from $t_{B2}$ and $t_{b3}$ connecting sites B and 
${\rm A^{\prime}}$. When the local distortion at site B making 
$t_{B2^{\prime}}>t_{B2}$ is weakened by photoexcitation, the hole is
locally transferred from site A through site B to site
A$^{\prime}$. This corresponds to the direction that weakens the
ferroelectric polarization. This results in the asymmetry of 
$\overline{|\Delta n_i|}$. It is in contrast to 
$\overline{|\Delta n_i|}$ for $\theta$-RbZn, where
the transfer integrals $t_{p}$ connecting sites A and C
are the same as $t_{p}$ connecting sites B and D. 

These numerical results suggest that a macroscopic domain
is much more easily created in $\alpha$-I$_3$ than in
$\theta$-RbZn. They are consistent with the experimental observations
by the time evolutions of reflectivity spectra\cite{Iwai_PRL07}, which
indicate the growth of a metallic domain only for $\alpha$-I$_3$,
although spatial inhomogeneity is not directly observed. The critical
slowing down observed in $\alpha$-I$_3$\cite{Iwai_PRL07} is beyond the
scope of the present study. If we consider it theoretically, we would
need to treat larger system and domain sizes. Effects of relaxation,
which are not taken into account here, may also be important for the
dynamics on longer timescales.

The different growth dynamics is caused by the different
mechanisms of stabilizing the COs: the
CO in $\theta$-RbZn is stabilized stripe by stripe through the
homogeneous modulation of transfer integrals along the stripes, while
the CO in $\alpha$-I$_3$ is stabilized locally, as explained in \S
2. The growth dynamics is thus qualitatively unchanged even if we use
different excitation frequencies and polarizations in the calculations.

\section{Summary}
We have investigated the photoinduced melting dynamics of COs in
quasi-two-dimensional organic conductors $\theta$-RbZn and
$\alpha$-I$_3$. 
Although they show COs with similar horizontal-stripe patterns, 
relative importance of e-ph couplings and their configurations of
transfer integrals in stabilizing the COs are quite different
\cite{Tanaka_JPSJ07,Tanaka_JPSJ08,Tanaka_JPSJ09ERRATA,Miyashita_PRB07,Miyashita_JPSJ08}.
By numerically solving the time-dependent Sch$\ddot{\rm o}$dinger
equation within the Hartree-Fock approximation, charge and lattice
dynamics are obtained in the extended Peierls-Hubbard models during and
after the oscillating electric field is introduced with and without
spatial dependence. 
We find different photoinduced dynamics in these salts, which originate
from different mechanisms of stabilizing the COs by lattice distortions.

In the case of a spatially uniform time-dependent electric field, we
calculated time-averaged hole densities as a function of the increment
in the total energy per site $\Delta E$. We find a critical value
$\Delta E_{c}$ above which the horizontal CO and the lattice
distortion simultaneously disappear. The $\Delta E_{c}$ value in $\theta$-RbZn is
larger than that in $\alpha$-I$_3$, as a consequence of the larger 
stabilization energy for the CO in $\theta$-RbZn.
When the COs are destroyed, the hole distribution reflects the symmetry
of the underlying crystal structure without lattice distortions. It
becomes uniform for $\theta$-RbZn whereas the
charge disproportionation between sites B and C remains for
$\alpha$-I$_3$. 

When the applied electric field is local, the time evolution of the hole
densities shows anisotropy. The influences of the local photoexcitation
propagate more rapidly to the direction parallel to the stripes than to that
perpendicular to the stripes. In particular, for $\theta$-RbZn a
photoinduced domain hardly expands to the perpendicular direction. 
The increase of $\Delta E/\Delta E_c$ is limited to a small value when
the pulse width is small, indicating that it is difficult to melt the CO
in $\theta$-RbZn by the local photoexcitation. 
This is because the $y_{\phi}$-distortion homogeneously modulates the
transfer integrals on the horizontally connected bonds, so that the
charge correlation in each stripe (between stripes) is strong
(weak). For
$\alpha$-I$_3$, on the other hand, the photoinduced domain expands in
the plane. This comes from the fact that each
hole-rich bond B2$^{\prime}$ between sites A and B in $\alpha$-I$_3$ is
locally stabilized by the $y_{b2}$-distortion.

The obtained results are qualitatively consistent with the experimental
observations, which indicate that the CO in $\theta$-RbZn is more stable
against photoexcitation than in $\alpha$-I$_3$: a macroscopic metallic
domain is generated in $\alpha$-I$_3$ whereas the CO only locally melts
in $\theta$-RbZn.

\section*{Acknowlegements}
The authors are grateful to S. Iwai and S. Miyashita for enlightening
discussions. This work was supported by Grants-in-Aid and ``Grand
Challenges in Next-Generation Integrated Nanoscience'' from the Ministry
of Education, Culture, Sports, Science and Technology of Japan.

\appendix
\section{Maximum of Increment in Total Energy}
	Figure 6 shows that $\Delta E/\Delta E_{c}$ begins to decrease at 
	around $\mid \mbox{\boldmath $E$}_\mathrm{ext}\mid =0.7\ (\mid
	\mbox{\boldmath $E$}_\mathrm{ext}\mid^2=0.5)$. 
	This behavior comes as a consequence of the periodicity of the
	energy band in momentum space. The electric field shifts the momenta
	of electrons. For simplicity, let us consider the
	noninteracting electrons in one-dimension, where the shift is
	estimated by the equation of motion,
\begin{equation}
	\frac{dk}{dt}=E_{\rm ext}\sin \omega_{\rm ext}t.
\end{equation}
	Here $k$ is the wave number and we set $e=a=\hbar =1$ with $a$
	being the lattice constant.
	By integrating the above equation from $t=0$ to
	$t=\pi/\omega_{\rm ext}$ that is one half of the period of the
	oscillating electric field, we obtain the momentum shift 
	$|\Delta k|$ as,
\begin{equation}
	|\Delta k|= \frac{2E_{\rm ext}}{\omega_{\rm ext}}.
\end{equation}
	As $E_{\rm ext}$ increases, $|\Delta k|$ and $\Delta E$ increase.
	When $|\Delta k|\sim \pi$, however, some electrons accelerated by the
	electric field cross the boundary of the Brillouin zone so that 
	$|\Delta k|$ is no longer proportional to $E_{\rm ext}$. If we increase 
	$E_{\rm ext}$ further, it is expected that the total energy is not
	efficiently increased any more and even decreases because more
	electrons cross the zone boundary. This results in the decrease
	of $\Delta E/\Delta E_{c}$ as shown in Fig. 6. 
	For $\omega_{\rm ext}=0.4$, $|\Delta k|\sim \pi$ corresponds to
	$E_{\rm ext}\sim 0.6$, which is consistent with the above value,
	$\mid \mbox{\boldmath $E$}_\mathrm{ext}\mid =0.7$.


\begin{thebibliography}{99}
\bibitem{Footnote}
	For $\alpha$-I$_3$, we use the symbol A$^{\prime}$ in stead of D which is
	used in ref. 32.
\end{thebibliography}
\end{document}